\documentstyle[preprint,aps]{revtex}
\begin{document}
\title{{\bf In--Medium Chiral Perturbation Theory and Pion Weak Decay 
in the Presence of Background Matter}} 
\author{M.\ KIRCHBACH$^1$ and A.\ WIRZBA$^2$}
\address{ $^1$Institut 
f\"ur Kernphysik, J. Gutenberg Universit\"at, D--55099 Mainz, Germany\\
$^2$Institut f\"ur Kernphysik, TH Darmstadt, 
D-64289 Darmstadt, Germany\\}
\maketitle
\begin{abstract}
Two--point functions related to the pion weak decay constant $f_\pi$
are calculated from the generating functional of chiral perturbation
theory in the mean--field approximation and the heavy--baryon
limit. The aim is to demonstrate that Lorentz invariance is violated
in the presence of background matter. This fact manifests itself in
the splitting of both $f_\pi$ and the pion mass into uncorrelated
time-- and spacelike parts. We emphasize the different in--medium
renormalizations of the correlation functions, show the inequivalence
between the in--medium values of $f_\pi$ deduced from Walecka--type
models, on the one hand, and QCD sum rules, on the other hand, and
elaborate on the importance for some nuclear physics observables.
\end{abstract}

\section{Introduction}

The symmetry underlying a physical system is viewed as one of the
basic guidelines in constructing the relevant dynamics.  The
determination of the conditions under which a symmetry is still valid
is therefore a question of fundamental interest.  Two--flavor chiral
symmetry ($\chi $S) (combined electric charge and parity independence
of strong interaction) has been suggested as the global internal
symmetry of the QCD lagrangian\cite{Colemen}.  As long as $\chi $S is
assumed to be realised in the non--multiplet Nambu--Goldstone mode
with pions acting as the associated Goldstone bosons, it becomes
possible to expand correlation functions in powers of the light quark
masses and the external pion momenta thought to be very small on the
hadronic scale of $\Lambda \sim 1\, GeV$.  More
specifically, within the framework of this {\em chiral perturbation
theory\/} ($\chi $PT) an effective low--energy generating functional
(denoted by $Z_{\rm eff}$) for Green's functions is developed under 
the assumption of chiral invariant couplings between the quark fields 
of the QCD action and the external sources (scalar ($s$), pseudoscalar
($p$), vector ($v_\mu$) and axial vector ($a_\mu$)).  The result is
that the source--extended action is even locally chiral symmetric
\cite{GL}.  Similarly, nucleons ($N (\bar N )$) are introduced through
their locally chiral invariant couplings to the external sources,
$\eta (\bar \eta $), respectively~\cite{GSS}.  As a consequence,
chiral QCD--Ward identities valid on the quark level are transcribed
in an unique way to the composite hadron level.  In the last decade
$\chi $PT advanced to an extremely powerful scheme for describing
various low--energy phenomena (see Ref.\,\cite{BKMreview} for a recent
review). In view of its successes the question on the applicability of
$\chi $PT in bulk matter~\cite{Leutwy} and
especially in
the case of nuclear matter at finite densities
\cite{KN,BLRT,ThWi} acquires actuality. We here investigate the
in--medium pion weak decay within the framework of $\chi $PT.  The
paper is organized as follows. In the next section we briefly review
the correlation function technique for introducing the pion weak decay
coupling constant $f_\pi$ at zero matter density.  In Sec.\,3 the
in--medium renormalization of the two--point functions is calculated
from the generating functional of $\chi $PT in the mean--field
approximation and the heavy baryon limit~\cite{MG,JM}.  We show that 
in the presence of background matter the various two--point functions
related to the pion weak decay matrix element have different density
dependences and therefore give rise to different definitions for the
in--medium $f_\pi$'s. We illustrate in Sec.\,4 the
implications for nuclear observables by calculating the
density dependence of the induced pseudoscalar coupling constant,
$g_p$. In combining the $\chi $PT results on the in--medium pion weak 
decay constants with QCD sum rules we also calculate the quark 
condensate dependence of the amplitude of the two--body axial charge 
density operator. The paper ends with a short summary.

\section{Defining $ f_\pi$ in the absence of background matter.
Two--point functions technique}

The pion weak decay constant is introduced through the matrix element
of the weak axial current $J_{\mu , 5 }^a = \bar q \gamma_\mu \gamma_5
{\tau^a\over 2} q$ between the pion and the hadronic vacuum state,
where $q$ denotes the isospin quark doublet.  According to
\cite{deAlf} the quantity $f_\pi$ can also be introduced via the
axial--current--pseudoscalar (AP) two--point function $T^{AP}_\mu$:
\begin{equation}
 T^{AP}_\mu  =  \frac{ i}{3} 
 \int {\rm d}^4x\, {\rm e}^{ik\cdot x}
 \langle 0|TJ_{\mu ,5}^a(x)\pi^a (0)|0\rangle = 
 f_\pi \, { 1 \over {-k^2 } }\, ik_\mu\, ,
 \label{Tap} 
\end{equation}
where $k_\mu $ is the pion four--momentum.
In the $k_0\to 0, \vec{k} =0$ limit its divergence reads
\begin{equation}
if_\pi = \lim_{\stackrel{\vec k=0 }{k_0\to 0}}
 -k^\mu T^{AP}_\mu
 =  
  \frac{1}{3}\int {\rm d}^3 x \,
 \langle 0|\lbrack J^a_{0, 5}(x), \pi^a (0)\rbrack |0\rangle
 =  \frac{1}{3}\langle 0|\lbrack Q^a_5(0), \pi^a(0)\rbrack|0\rangle \, .
 \label{AP}
\end{equation}
Here $Q_5^a(0)$ is the integrated axial charge density,
$J_{0, 5}^a (x)$.
The conserved axial current hypothesis (CAC) has
been exploited for simplicity.  In evaluating the commutator in
Eq.\,(\ref{AP}) within the linear $\sigma$ model where 
$\lbrack Q_5^a(0),\pi^b(0)\rbrack = -i\sigma \delta^{ab}$, one finds 
that $f_\pi $ equals the negative vacuum expectation value of the 
$\sigma$ field, $f_\pi = - \langle 0|\sigma|0\rangle $. Obviously, it 
is the $AP$ correlator that underlies the definition of $f_\pi $ within 
the linear $\sigma $ model. Consider now the correlator (subsequently 
denoted by $(AA)$) of two axial vector currents
\begin{equation} 
 T^{AA}_{\mu ,\nu} = {i\over 3} \int {\rm d}^4x\, {\rm e}^{ik\cdot x} 
 \langle 0| TJ^a_{\mu ,5}(x) J^a_{\nu , 5}(0)|0\rangle 
 = f_\pi^2 \, ( g_{\mu \nu }  +
 {{k_\mu k_\nu }\over {m_\pi^2 -k^2 }}) \, , 
\label{AA_corr}
\end{equation}
and the closely related Gell-Mann--Oakes--Renner (GOR) 
correlator \cite{GeOaRe}
\begin{equation}
 \lim_{\stackrel{\vec k=0}{k_0\to 0}}
 {i\over 3} \int {\rm d}^4x\, {\rm e}^{ik\cdot x} 
 \langle 0| T \partial ^\mu J^a_{\mu ,5}(x) \partial^\nu 
 J^a_{\nu , 5}(0)|0\rangle 
 = -f_\pi^2 m_\pi^2\, .
\label{AA_GOR}
\end{equation}
The l.h.s.\ of Eq.\,(\ref{AA_GOR}) can furthermore be evaluated
\cite{GeOaRe} to give $ {1\over 2} (m_u+m_d)\langle 0|\bar u
u\mbox{+}\bar d d|0\rangle $. As a result $f_\pi$ can be 
now extracted from
\begin{equation}
   f_\pi^2 m_\pi^2 = -{{m_u\mbox{+}m_d}\over 2} 
 \langle 0|\bar u u\mbox{+}\bar d d|0\rangle \,  
 \equiv - 2m_q \langle 0|\bar q q |0\rangle \, ,
 \label{GOR_1}
\end{equation}
where we used the approximation 
$\langle 0|\bar u u|0\rangle = \langle 0|\bar d d|0\rangle
\equiv \langle 0|\bar q q|0\rangle $. 
Eqs.\,(\ref{AP}), (\ref{AA_corr}), and (\ref{AA_GOR}) define the
vacuum  $f_\pi$ in terms of different two--point
functions at tree-level order. As we will see in the next section, 
this is not possible in the case of matter at finite densities, as the
various correlation functions will acquire different in--medium
dependences.

\section{In--medium two--point functions for the pion weak decay 
constant from the generating functional of $\chi $PT in the 
mean--field approximation}

The generating functional of $\chi $PT is written as usual according
to
\begin{equation}
 {\rm e}^{i Z_{\rm eff}\lbrack s,p, v_\mu ,a_\mu,\eta , \bar
 \eta \,\rbrack}
 ={\cal N} \int {\rm d}U{\rm d}N{\rm d}\bar N 
 {\rm e}^{i\int {\rm d}^4x\, L_{\rm eff}}\; ,
 \label{Gen_Fct}
\end{equation}
where ${\cal N}$ is an overall normalisation constant, $U,N$ and 
$\bar N$ are dummy boson and fermion fields and the effective 
lagrangian $L_{\rm eff}$ depends on possible external sources, 
$s,p, v_\mu ,a_\mu,\eta, \bar \eta$. Our goal now is to calculate 
the in--medium renormalization
of the two--point functions (\ref{AP}), (\ref{AA_corr}), and
(\ref{AA_GOR}) defined in the previous section.  We start with the
effective chiral pion-nucleon lagrangian to order ${\cal O}(Q^2)$ in 
the heavy baryon limit~\cite{MG,JM}. This lagrangian is relevant for 
the isoscalar low--energy $\pi N$ scattering 
(see Refs.\,\cite{BKM,BLRT,ThWi} for more details)
\begin{eqnarray}
 L_{\rm eff}  &=& i\bar N (v\cdot \partial -\sigma_N) N 
 +{1\over 2}(\partial _\mu \pi)^2 
 -{1\over 2}m_\pi^2\pi^2 \nonumber \\
 && \mbox{}+  {1\over f_\pi^2}
 \left \{ {1\over 2}\sigma_N \pi^2 +c_2(v\cdot \partial \pi)^2
  +c_3 (\partial_\mu \pi)^2 \right \}\bar N N
 + j^a\pi^a (1-{{\sigma_N \bar N N}\over {f_\pi^2m_\pi^2}})
 + \cdots\, .
 \label{eff_La}
\end{eqnarray}
Here, $v_\mu$ is the 4--velocity of the nucleon ($N$) in the heavy
baryon limit and $\pi^a$ stands for the pion field at zero matter
density. The pseudoscalar source has been rescaled to $j^a \equiv
2Bf_\pi p^a$, where $B=-m_\pi^2 /(m_u\mbox{+}m_d)$.  The combination
$(c_2+c_3)m_\pi^2$, if extracted on the tree level from the empirical
isospin--even pion nucleon scattering length $a^+_{\pi N}$, takes
the value $(c_2+c_3)m_\pi^2 = -26\,{\rm MeV}$ \cite{BKM}, while
$\sigma_N$, at order order ${\cal O}(Q^2)$, stands for the
pion--nucleon sigma term with $\sigma_N = 45\,{\rm MeV}$.  The
presence of background matter is taken into account by the replacement
of $\bar N N$ in $L_{\rm eff}$ with the matter density $\rho$. Using
the quaternion notation $U = \exp(i\tau^a\pi^a/f_\pi)$ the lagrangian
can be cast into the mean--field form
\begin{equation}
 {L}_{\rm MF} = {f_\pi^2\over
 4}(g^{\mu \nu } +{{D^{\mu\nu }\rho}\over f_\pi^2}) Tr((\nabla_\mu
 U)^\dagger\nabla _\nu U) +{f_\pi^2\over 4} (1- {{\sigma_N\rho }\over
 {f_\pi^2m_\pi^2}}) Tr (U^\dagger \chi +\chi^\dagger U) + ... \, , 
 \label{mean_La}
\end{equation}
where $D^{\mu\nu} \mbox{$\equiv$} 2c_2 v^\mu
v^\nu\mbox{+}2c_3g^{\mu\nu}$, $\nabla_\mu U \mbox{$\equiv$}
\partial_\mu U\mbox{$-$}i{1\over 2} \{\tau^a a^a_\mu,U\} \mbox{$-$}i{1
\over 2} [\tau^a v^a_\mu, U ]$ and $\chi \mbox{$\equiv$}
2B(s\mbox{+}i\tau^a \pi^a)$.  Within this $\chi $PT scheme, the
calculation of the in--medium two--point functions is now a
well--defined procedure, since the Green's functions are obtained by
taking functional derivatives of the corresponding generating
functional $Z_{\rm MF}[s,p,v,a;\rho]$ with respect to the sources:
\begin{eqnarray}
 T_\mu^{AP} &=& 
 \frac{1}{3}\,{{\delta^2 Z_{\rm MF}}\over 
 {\delta a^{a\,\mu}(-k)\delta p^ a (k)}}|_{a=v=p=0, s={\cal M}}\, , \\
  T_{\mu ,\nu}^{AA} &=&\frac{1}{3}\,
 {{\delta^2 Z_{\rm MF}}\over 
 {\delta a^{a\,\mu}(-k)\delta a^{a\,\nu} (k)}}|_{a=v=p=0, s={\cal M}}\, 
  , \\
 T^{PP}& =&\frac{1}{3}\,
 {{\delta^2 Z_{\rm MF}}\over {\delta p^a (-k)\delta p^ a (k)}}
 |_{a=v=p=0, s={\cal M}}\, .
 \label{func_der}
\end{eqnarray}
The latter correlation function is associated with the pion
propagator.  In Refs.\cite{ThWi,WiTh,KiWi} different aspects of its
in--medium version were considered and the following expressions were
reported at the tree-level order
\begin{eqnarray}
 \frac{i}{3}\int {\rm d}^4x\, 
 {\rm e}^{ik\cdot x}\langle \tilde{0}|T P^a(x) P^a(0)|\tilde{0}
 \rangle &=&
 -{{G_\pi^*\, ^2}\over 
 {k_0^2 -\gamma (\rho)\vec{k\, }^2 -m_\pi^*\,^2}}\, ,
 \nonumber\\
 G_\pi^*\, ^2 =G_\pi^2 
 {{(1-{{\sigma_N\rho }\over {f_\pi^2m_\pi^2}})^2}\over 
 {1 +{{2(c_2+c_3)\rho}\over f_\pi^2}}} \, , &\quad &
 c_2\mbox{+}c_3 = -0.27 \,{\rm fm}\, 
 ,\quad  c_3 = -0.55\,{\rm fm} \, 
 ,\nonumber\\
 m_\pi^* \, ^2 =
 { {1- {{\sigma_N\rho}\over {f_\pi^2m_\pi^2}} }\over
 {1 +{{2(c_2+c_3)\rho }\over f_\pi^2}} }m_\pi^2 \, ,
 &\quad &
 \gamma (\rho ) = {{1 +{{2c_3\rho}\over f_\pi^2}}\over
 {1 +{{2(c_2+c_3)\rho}\over f_\pi^2}}}\, , \quad \sigma_N = 0.228\, 
 {\rm fm}^{-1}\, .
 \label{par_numb}
\end{eqnarray}
Here $G_\pi$ is the overlap of the pseudoscalar
density $P^a\mbox{$\equiv$}\bar q i \gamma_5 \tau^a q$
between the vacuum and the pion state:
$G_\pi \delta^{ab}\mbox{$\equiv$}\langle 0|\bar q i\gamma_5\tau^a
q|\pi^b \rangle$. Therefore $j^a = G_\pi p^a$. 

The pion field corresponding to the propagator in
Eq.\,(\ref{par_numb}) with weight one for the combined $k_0^2$ term is
the so--called {\em quasi--pion} field $\widetilde{\phi }^a_\pi$. On
the other hand, the commonly used {\em Migdal
field\/}\,\cite{KiRi,KiWi} (denoted by $\widetilde{\pi}^a$)
corresponds to the case when the inverse propagator has the form $k^2
-m_\pi^2 -\Pi (k_0, \vec{k}\, )$ with $\Pi (k_0, \vec{k}\, )$ standing
for the pion self energy. Explicit expressions for $\Pi (k_0,
\vec{k})$ can be found in Ref.  \cite{KiWi}.  The quasi--pion and the
Migdal field are related to the bare pion field by
\begin{equation}
 \widetilde{\phi }_\pi^a =
 {G_\pi \over G_\pi^*} \widetilde{\pi} ^a\, , \qquad 
 \widetilde{\pi}^a =  
 {{\delta Z_{\rm MF}\lbrack j^a,\rho \rbrack } \over {\delta j^a}}
 = \left(1-
 {{\sigma_N\rho }\over {f_\pi^2m_\pi^2}}\right)\,\pi^a\, ,
 \label{pi_fields}
\end{equation}
respectively. Note that ${m_\pi^*}^2$ from Eq.\,(\ref{par_numb})
corresponds to the pole of the in--medium pion propagator with respect
to a purely timelike pion momentum. At normal nuclear matter density
($\rho_0 = 0.17\, {\rm fm}^{-3}$) $m_\pi^*\,^ 2 $ almost equals the
free pion mass $m_\pi^2$. If we had considered a purely spacelike pion
momentum there, the corresponding mass--pole would be placed at
\begin{equation}
 {m_\pi^*}^2 = -
 m_\pi^2 {{1 - \frac{\sigma_N\rho_0}{f_\pi^2m_\pi^2} } \over
 {1+ \frac{2c_3\rho_0}{f_\pi^2}}} +{\cal O}(m_\pi^3)\approx -4m_\pi^2\, .  
 \label{P_pole}
\end{equation}
The difference in the absolute values of the timelike and spacelike 
mass--poles of the in--medium pion propagator signals a violation of 
Poincar\'{e} invariance in the presence of background matter and 
indicates that the in--medium hadronic states are no longer eigenstates 
of the ${\cal P}(1,3)$ Casimir operator $k^\mu k_\mu $.  Similarly, 
the currents which we have considered do not transform any longer 
according to the $\lbrace {1\over 2}, {1\over 2}\rbrace $ representation 
of the Lorentz group, but decompose after its contraction to the Galilei 
group into charge and current parts\,\cite{KiRi}. The timelike in--medium
$AP$--correlator evaluated along the $\chi $PT\, line \cite{WiTh,KiWi}
at the tree-level order is
\begin{equation}
 T_0^{AP}\, =\, 
 \frac{i}{3}\int {\rm d}^4 x\, {\rm e}^{ik\cdot x}
 \langle \tilde{0}|TJ_{ 0 ,5}^a (x) P^a (0)|\tilde{0}\rangle 
 = f_\pi \left(1 - {{\sigma_N\rho }\over {f_\pi^2m_\pi^2}}\right )\, 
 {G_\pi
 \over {m_\pi^* \, ^2 -k^2_0 +\gamma (\rho) \vec {k\, }^2 }}\,ik_0 .
 \label{AP_1}
\end{equation}
Note that we use here the axial-current-pseudoscalar 
two-point function of  $\chi PT$ which is formulated in
terms of the pseudoscalar density $P^a$, because it is this quantity 
that couples directly to the external pseudoscalar source 
$p^a$~\cite{GL}. Thus, even in the normal vacuum, this correlator is 
multiplied by a factor $G_\pi$  relative to the one of Eq.~(\ref{Tap}) 
which is just expressed in terms of  the pion field $\pi^a$ instead. 
Taking this factor into account, the comparision of (\ref{AP_1}) and 
(\ref{Tap}) shows that the in--medium pion weak decay constant 
associated with $T^{AP}_\mu $ is now given by
\begin{equation}
 f_\pi^{AP}(\rho )
 =\lim_{{\vec{k}}\to 0;{m_\pi^\ast}^2\to 0}\, \frac{i k_\mu}{3 G_\pi}\,
 {{\delta^2 Z_{\rm MF}\over 
 {\delta a_\mu^a(-k)\, \delta p^a(k) }}}|_{a_\mu=v_\mu=p=0,s={\cal M}}
 = f_\pi\, \left(1 -{{\sigma_N\rho }
 \over {f_\pi^2m_\pi^2}}\right )\, .
 \label{fpi_AP}
\end{equation}
This exactly reproduces the in--medium expectation value of the
$\sigma$ field in Walecka--type models, as it should, since Walecka
type models have their roots in the linear $\sigma$ model. Finally,
the in--medium $T^{AA}_{00}$ is evaluated as\,\cite{ThWi}
\begin{equation}
 T^{AA}_{00}(\rho\mbox{$\not=$}0) 
 = f_\pi^2 \left(1 + {{2(c_2+c_3)\rho }\over f_\pi^2}\right)
  \left(g_{00}+ 
 { k_0^2 \over {m_\pi^*\, ^2 -k_0^2 +\gamma (\rho ) \vec {k\, }^2}} 
 \right) \, .
 \label{AA_1}
\end{equation}
{}From that, in combination with Eq.\,(\ref{GOR_1}) and the 
in--medium version of (\ref{AA_GOR}), the in--medium GOR 
relation becomes~\cite{KiWi}
\begin{equation}
  -2m_q
 \langle \tilde{0}|\bar q q|\tilde{0}\rangle
 = f_\pi^2 m_\pi^2 
 \left(1\mbox{$-$}{{\sigma_N\rho }\over {f_\pi^2 m_\pi^2}}\right)\,
 ={f_\pi^{(t)}}^2 {m_\pi^*}^2 \,  
 , \quad
 {f_\pi^{(t)}}^2 
 = f_\pi^ 2\,  \left(1\mbox{+}{{2(c_2\mbox{+}c_3)\rho }\over f_\pi^2} 
 \right)\, .
\label{med_GOR}
\end{equation}
In fact, the in--medium pion decay constant $f_\pi^{(t)}$ {\em
associated with the GOR relation} is independent of the choice for 
the in--medium pion field\,\cite{ThWi}. On the other hand, 
the in--medium weak pion decay constants {\em entering the 
time/space like matrix elements} between
the pion and the dense vacuum states (which describe
in--medium $S$-- and $P$--wave pion weak decays, respectively) depend 
on this choice. They can be directly read
off from the kinetic part of the mean--field lagrangian in
Eq.~(\ref{mean_La}), more precisely, from the coupling of the external
axial vector source $a^a_\mu$ to the axial vector current expressed by
the derivative of the in--medium
pion.  Depending on the choice for the pion field (quasi--pion or
Migdal's), the following expressions hold at normal nuclear matter
density $\rho_0$ to tree-level order:
\begin{eqnarray}
 \langle \tilde{0}|J_{\mu ,5}^a|\tilde{\pi}^a\rangle 
 = (i f_\pi^S\, k_0\,,\, i f_\pi^P \, \vec{k}\, ), &\qquad &
 f_\pi^S =f_\pi
 { {1 +{ {2(c_2+c_3)\rho_0 }\over f_\pi^2}}\over
 {1- { {\sigma_N\rho_0}\over {f_\pi^2m_\pi^2}}}}   
 \approx 0.9f_\pi\, ,\nonumber\\ 
 \label{ma_el1}
 f_\pi^P\! =\!f_\pi
 {{1{+}{{2c_3\rho_0}\over f_\pi^2}}\over 
 {1{-}{ {\sigma_N\rho_0}\over {f_\pi^2m_\pi^2} }}}
 \!\approx\! {1\over 4}f_\pi \, , &\qquad &
 \langle \tilde{0}|J^a_0 |\tilde{\phi }_\pi^a \rangle \!=\!
 f_\pi\sqrt{1\mbox{+}{{2(c_2\mbox{+}c_3)\rho_0}\over f_\pi^2}} ik_0
 \equiv f_\pi^{(t)}ik_0\, ,\\
 f_\pi^{(t)}  = {G_\pi^*\over G_\pi} f_\pi^S 
=f_\pi \sqrt{1+{{2(c_2+c_3)\rho_0}\over f_\pi^2}}\, ,
&&
f_\pi^{(s)}={G_\pi^*\over G_\pi}f_\pi^P 
= f_\pi\,  {{1 +{{2c_3\rho_0}\over f_\pi^2}}\over
{\sqrt{1+{{2(c_2+c_3)\rho_0}\over f_\pi^2}}}}\, .
 \label{ma_el}
\end{eqnarray}
These equations reveal the fundamental difference between
$f_\pi^{AP}$, on the one side, and $f_\pi^{(t)}$, on the other side.
Whereas $f_\pi^{AP}$ cannot be directly attributed to a physical
process, $f_\pi^{(t)}$ parametrizes the weak decay matrix element of
an $S$--wave quasi--pion.  Therefore, while $f_\pi^{(t)}$ is directly
related to an observable, $f_\pi^{AP}$ should be viewed as an
unphysical quantity.

\section{Density dependence of the induced pseudoscalar coupling 
and the amplitude of the two--body axial charge density}

The in--medium renormalized pion weak decay constants calculated above
enter various nuclear physics observables and can in principal be
extracted from nuclear physics measurements.  As a first example we
consider here the in--medium renormalization of the induced
pseudoscalar coupling strength $g_p (k^2)$ \cite{KiRi}
\begin{equation}
g_p (k^2) = -2m_\mu {  {f_\pi g_{\pi NN}}\over {k^2 - m_\pi^2}}\, ,
\label{g_P}
\end{equation}
where $m_\mu $ stands for the muon mass and $g_{\pi NN}$ denotes the 
pseudoscalar
pion nucleon coupling constant. 

A systematical evaluation of $g_p(k^2)$ at finite nuclear matter
density would require knowledge of the in--medium behavior of the
strong pion--nucleon vertex.  As long as at that stage such knowledge
is still absent, we will exploit here QCD sum rules to argue that the
$\rho $ dependence of $g_{\pi NN} $ can be ignored for the time being.
To get the dependence of $g_{\pi NN}$ on $\langle 0|\bar q q|0\rangle
$ in the vacuum, we use the estimates from QCD sum rules as evaluated
in Refs.~ \cite{Reinders},\cite{Ioffe}:
\begin{equation} 
 {g_{\pi NN}^2\over {4\pi }} \approx 2^5\pi^3 {{f_\pi^2 }/ m_N^2}\, ,
\qquad  m_N^3  \approx -8\pi^2 \langle 0|\bar q q|0\rangle \, .
\label{QCD_coupl}
\end{equation}
In fact, for $\langle 0|\bar q q|0\rangle\mbox{=}-1.48\,{\rm
fm}^{-3}$, the estimates given in Eq.~(\ref{QCD_coupl}) are easily 
verified to be in good agreement 
with the corresponding experimental values. 
These estimates basically  imply that the quark condensate is in the 
vacuum the only  parameter setting the mass scale 
in the relevant QCD sum rules.
In particular,  the QCD sum rule for $f_\pi^2$ obtained 
from  the correlator of two axial currents   
\cite{svz} (see also \cite{Radyu}) is dominated 
by  the quark condensate and, hence,  
$f_\pi^2 \sim \langle 0|\bar q q|0\rangle ^{2/3} $. 
As a result, 
$g_{\pi NN}$ can be viewed as almost independent of the
quark condensate. 
This property of the pion nucleon coupling constant is most clearly 
seen when the ratio ${f_\pi^2\over m_N^2}$ in the QCD sum rule for 
$g_{\pi NN}^2 $ is replaced by the equivalent ratio 
${g_A^2\over g_{\pi NN}^2}$ resulting from the Goldberger--Treiman 
relation, where $g_A$ denotes the isovector weak axial nucleon 
coupling. In doing so one obtains the remarkable relation 
\begin{equation}
g_{\pi NN} \approx 4\pi \, \sqrt{g_A\over \sqrt{ {2} } }  
\approx 4\pi\, .
\label{gpiNN_fund}
\end{equation}
Because of that the in--medium change of the induced pseudoscalar
coupling will be subsequently ascribed exclusively to the 
renormalisation of $f_\pi$ as well as of the pion propagator.

Eqs.\,(\ref{ma_el1}) and 
(\ref{ma_el}) show that it is important to consider the correct 
combination of the weak pion decay constant and the pion propagator. 
In other words, Migdal's propagator has to be combined in turn with 
$f_\pi^P$ or $f_\pi^S$ to obtain the induced pseudoscalar couplings
corresponding to a P--wave ( $g_p^{(P)}(k_0^2,\vec{k}^2; \rho )$),
and an S--wave ($g_p^{(S)} (k_0^2,\vec{k}^2;\rho $)) in--medium 
pion, respectively. In doing so, $g_p^{(S)}(k_0^2,\vec{k}^2; \rho ) $ 
emerges from
\begin{equation}
g_p^{(S)}(k_0^2, \vec{k}^2; \rho ) = -2m_\mu 
{  {f_\pi^S g_{\pi NN}}\over 
{ \alpha (\rho ) k_0^2- \beta (\rho ) \vec{k}^2 - \theta (\rho ) m_\pi^2} } 
\label{gp_S1}\, .
\end{equation}
The coefficients in front of $k_0^2$, $\vec{k}^2$,
and $m_\pi^2$ have been determined in \cite{KiWi}. They read: 
\begin{eqnarray}
\alpha (\rho ) = 
{ 
{1+ { {2(c_2+c_3)\rho }\over f_\pi^2} } \over
{ (1-{ {\sigma_N\rho}\over {f_\pi^2m_\pi^2} })^2}  } \, , && 
\beta (\rho ) = { 
{ 1+ { {2c_3\rho}\over f_\pi^2} }\over
{(1-{ {\sigma_N\rho}\over {f_\pi^2m_\pi^2}})^2} }\nonumber\\
\theta (\rho ) &=& {1\over {1 - {{\sigma_N\rho }\over 
{f_\pi^2m_\pi^2}}}}\,
\label{mig_prop}
\end{eqnarray}
If the the quasi--pion propagator is used instead of Migdal's
propagator, $f_\pi^S $ and $f_\pi^P$ have to be rescaled to 
${G_\pi^*\over G_\pi}f_\pi^S$ , and $ {G_\pi^*\over G_\pi}f_\pi^P$,
respectively. The timelike and spacelike parts of 
the in--medium induced pseudoscalar coupling generated in this way
will subsequently be denoted in turn by
$g_p^{(t)}(k_0^2,\vec{k}^2;\rho )$ and $g_p^{(s)}(k_0^2,\vec{k}^2;\rho )$.
For example, for $g_p^{(t)}(k_0^2,\vec{k}^2;\rho )$ one has   
\begin{equation}
g_p^{(t)} (k_0^2, \vec{k}^2 ;\rho ) = -2m_\mu 
{{f_\pi^{(t)}g_{\pi NN}}\over 
{k_0^2 -\gamma (\rho )\vec{k}^2 -m_\pi^*\, ^2}}
\label{gp_S2}
\end{equation}
with $\gamma (\rho )$ being defined in Eq.~(\ref{par_numb}).
With that, the ratio $g_p^{(t)}(k_0^2,\vec{k}^2;\rho) /g_p(k^2) $ 
is predicted as
\begin{equation}
{ {g_p^{(t)} (k_0^2, \vec{k}^2;\rho )}\over {g_p (k^2)}}=
{ f_\pi^{(t)}\over f_\pi }
{ {k^2 - m_\pi^2}\over
{k_0^2 -\gamma (\rho) \vec{k}^2 -m_\pi^*\, ^2} }\, .
\label{gps_gp}
\end{equation} 
Within this scheme the $g_p^{(s)}(k_0^2,\vec{k}^2;\rho ) /g_p(k^2) $ 
ratio is now calculated as
\begin{equation}
{{g_p^{(s)} (k_0^2;\vec{k}^2 ; \rho ) }\over {g_p (k^2)}}
= {G_\pi^*\over G_\pi} \, {f_\pi^P \over f_\pi}
\, {{k^2 - m_\pi^2}\over
{k_0^2 -\gamma (\rho) \vec{k}^2 -m_\pi^*\, ^2}}\, .
\label{gpP_gp}
\end{equation}
Numerical evaluation of Eqs.~(\ref{gps_gp}) and (\ref{gpP_gp}) for the
case of, say, radiative muon capture on $^{40}$Ca with
$|k_0| = 0.8 m_\mu , |\vec{k}| = 0.2 m_\mu $, and $\rho = 0.8\rho_0$, 
leads to the renormalizations
${{g_p^{(t)} (k_0^2;\vec{k}^2 ; \rho ) }\over {g_p (k^2 )}}
\approx 0.73$, and
${{g_p^{(s)} (k_0^2;\vec{k}^2 ; \rho ) }\over {g_p (k^2 )}}
\approx 0.37$, respectively.
The suppression with density of both 
$g_p^{(s)}(k^2_0;\vec{k}^2;\rho )$ and 
$g_p^{(t)}(k^2_0;\vec{k}^2;\rho )$  
relative the vacuum value of the induced pseudoscalar coupling
$g_p (k^2)$ is supported by data through
a decreasing effective $g_p/g_A$ ratio in radiative muon capture 
reactions when going from $^3$He to $^{206}$Pb \cite{Harold}, 
\cite{Hass}. Note,  that our  prediction for the density dependence 
of the induced pseudoscalar coupling of a $P$--wave in--medium pion 
is less reliable than that for $g_p^{(t)}(k_0^2,\vec{k}^2; \rho ) $, 
because of a large uncertainty in the value of the parameter $c_3$ 
(see \cite{BKM} for a discussion) that controls the in--medium 
behavior of $f_\pi^{(P)}$ as presented in Eq.~(\ref{ma_el}).  

\noindent
Let us consider as a second example  the amplitude of
the two--body $(2b)$ axial charge operator 
\begin{equation}
 J_{0,5}^a (2b) = C_{\rm MEC}^{\rho =0}
                   \sum_{i\not=j}(\vec{\sigma}_i 
 +\vec{\sigma}_j)\cdot \hat{r}_{ij}
                    Y_1(m_\pi r)(\vec{\tau_i}
 \times \vec{\tau_j})^a\, ,\quad
 C_{\rm MEC}^{\rho =0 } =  
 { {g_{\pi NN}^2\, m_\pi^2} \over {8\pi g_A m_N^2}}\, .
 \label{MEC}
\end{equation}
Here, $Y_1(x)$ denotes the first order Yukawa function, 
$Y_1 (x)= {e^{-x}\over x}(1+{1\over x})$.
An appealing possibility to calculate  
$C_{\rm MEC}^{\rho }$ for $\rho \neq 0$
is to combine our $\chi $PT results on the in--medium 
weak pion decay coupling constants 
with the vacuum predictions of the QCD sum rules
and to assume that the $\rho$-dependence
of $C_{\rm MEC }^\rho $ is generated by the change of the 
quark condensate value in the nuclear medium under the constraint 
of the in--medium  GOR relation.
In making use of Eq.~(\ref{QCD_coupl}) and the vacuum GOR relation 
(\ref{AA_GOR}), which according to  Ref.\cite{svz} is completely 
consistent with the QCD sum rule approach, one finds
the following expression for $C^{\rho =0 }_{\rm MEC}$:
\begin{equation}
C^{\rho =0 }_{\rm MEC} \approx 2^4\pi^3 
{ {f_\pi^2 m_\pi^2}\over { g_A m_N^4}}\, 
= 2^4\pi^3 
{ {-2m_q \langle 0|\bar q q|0\rangle }\over 
{g_A
(-8\pi^2\langle 0|\bar q q|0\rangle )^{4\over 3} }
} \, . 
\label{QCD_MEC}
\end{equation}
Here,  we have ignored the quark condensate dependence of $g_A$.
Indeed, according to the QCD sum rule for $g_A$ reported
in Ref.\,\cite{BelKog} it is only the small deviation of $g_A$ from 
unity that depends on the quark condensate in accordance with
\begin{equation}
 g_A-1 \approx 0.13\,{\rm fm}^6\,\langle 0|\bar q q|0\rangle ^2 \, .
\end{equation}
{}Furthermore, our assumption is that $C^{\rho }_{\rm MEC}$ in nuclear 
matter follows from this vacuum relation, if the vacuum value of the 
quark condensate is replaced by its in-medium value in accordance with 
Eq.\,(\ref{med_GOR}). We thus obtain the ratio 
$C^{\rho =\rho_0}_{\rm MEC} /C^{\rho =0}_{\rm MEC}$ as,
\begin{equation}
 {C_{\rm MEC}^{\rho=\rho_0} \over
 C_{\rm MEC}^{\rho =0}} \approx 
 \left( 
 { {\langle 0| \bar q q|0\rangle }\over
  {\langle \tilde{0}| \bar q q| \tilde{0} \rangle } 
    }\,\right )^{{1\over 3}} 
  = \left(1 - {{\sigma_N\rho_0}\over
 {m_\pi^2f_\pi^2}}\right)^{-{1\over 3}} \approx 1.15 \, .
\label{MEC_renorm} 
\end{equation}
The result in Eq.\,(\ref{MEC_renorm}) shows that the
$40\%$ enhancement of $C_{\rm MEC}^\rho$ (needed to explain the
acceleration of isovector first--forbidden weak decays in the lead
region \cite{Ernie}) is not completely exhausted by renormalization
effects due to the presence of background matter.  
There is still room left for contributions of short range heavy meson 
exchanges to the two--body axial charge operator, in line with 
the suggestions considered in\, \cite{KiRiTsu} and \cite{Towner}.

\section{Summary}

In calculating in--medium properties of pion weak decay within the
framework of $\chi $PT in the mean field approximation and
the heavy baryon limit,
we demonstrated a violation of Poincar\'{e}
invariance in the presence of background matter. This fact manifests
itself in the splitting of both $f_\pi$ and the pion mass (considered
as a pole of the in--medium pion propagator) into uncorrelated
timelike and spacelike parts.  A further need for distinguishing 
between timelike and spacelike pion weak decay matrix elements emerges
in case of nuclear matter at finite temperature, an observation 
reported in Ref.~\cite{MTyt}. We furthermore revealed the unphysial 
nature of the in--medium expectation value of the $\sigma $ field
associated with the pion weak decay constant in hadrodynamical models 
of Walecka--type.
This contrasts the model independent quantity $f_\pi^{(t)}$ that 
determines the in--medium weak decay matrix element of
an $S$--wave quasi--pion and enters the GOR relation.
We further investigated the in--medium timelike and spacelike 
induced pseudoscalar couplings $g_p^{(t)} (k_0^2,\vec{k}^2;\rho )$,
and $g_p^{(s)} (k_0^2,\vec{k}^2;\rho )$. We found them both 
suppressed relative to their values at zero matter density,
in line with experimental observations in radiative
muon capture reactions.
Finally, we studied the dependence of 
the amplitude $C_{\rm MEC}^{\rho }$ of the two--body axial charge 
density operator $J_{0,5}^a(2b)$  on the in--medium quark condensate
in combining our $\chi $PT results on the in--medium
pion weak decay matrix elements with QCD sum rules..
We showed that there still is some room left for
contributions to $J_{0, 5}^a (2b)$ due to short--range 
heavy meson exchanges between the nucleons.

To close, we wish to stress once more that the concept of
in--medium $\chi $PT represents a mathematically well defined scheme 
which allows a consistent evaluation of various hadron couplings 
measured in nuclear physics processes. 

\section{Acknowledgements}

This  work  was partly supported by Deutsche Forschungsgemeinschaft 
(SFB 201). One of us (M.K.) likes to thank Anatoly Radyushkin for 
hospitality at TJNAF as well as for numerous enlightening discussions
on various aspects of QCD sum rules.

\end{document}